\documentclass[authoryear,final,5p,times,twocolumn]{elsarticle}
\def\astrobj#1{#1}

\usepackage{graphicx,rotating}

\usepackage{amssymb}

\journal{New Astronomy}

\begin{document}

\begin{frontmatter}


\title{An {\it XMM-Newton} view of the M\,17 nebula\tnoteref{label1}}
\tnotetext[label1]{Based on observations collected with {\it XMM-Newton}, an ESA Science Mission with instruments and contributions directly funded by ESA Member States and the USA (NASA).}

\author{F.\ Mernier}
\ead{mernier@astro.ulg.ac.be}

\author{G. Rauw}
\ead{rauw@astro.ulg.ac.be}

\address{Groupe d'Astrophysique des Hautes Energies, Institut d'Astrophysique et de G\'eophysique, Universit\'e de Li\`ege, 17, All\'ee du 6 Ao\^ut, B5c, B-4000 Sart Tilman, Belgium}


\begin{abstract}
We present the analysis of an {\it XMM-Newton} observation of the M\,17 nebula. The X-ray point source population consists of massive O-type stars and a population of probable low-mass pre-main sequence stars. CEN\,1a,b and OI\,352, the X-ray brightest O-type stars in M\,17, display hard spectra (kT of 3.8 and 2.6\,keV) consistent with a colliding wind origin in binary/multiple systems. We show that the strong interstellar reddening towards the O-type stars of M\,17 yields huge uncertainties on their $L_{\rm X}/L_{\rm bol}$ values. The low-mass pre-main sequence stars exhibit hard spectra resulting from a combination of high plasma temperatures and very large interstellar absorption. We find evidence for considerable long term (months to years) variability of these sources. M\,17 is one of the few star formation complexes in our Galaxy producing diffuse X-ray emission. We analyze the spectrum of this emission and compare it with previous studies. Finally, we discuss the Optical Monitor UV data obtained simultaneously with the X-ray images. We find very little correspondence between the UV and X-ray sources, indicating that the majority of the UV sources are foreground stars, whilst the bulk of the X-ray sources are deeply embedded in the M\,17 complex. 
\end{abstract}


\end{frontmatter}


\section{Introduction \label{intro}}
\astrobj{M\,17} is one of the most luminous H\,{\sc ii} regions of our Galaxy. This nebula hosts the very young open cluster \astrobj{NGC\,6618} which probably reflects the second stage in a still on-going sequential (massive) star formation process \citep{Hanson,Povich09}. These properties make M\,17 a privileged target for the observational study of massive star formation. Indeed, M\,17 harbors several cocoon stars, which are still embedded in the relics of their cradles of circumstellar material \citep[e.g.][]{CK,Hanson,Chini,Kassis}. Several interesting objects are found along the interface between the H\,{\sc ii} region and the molecular cloud M\,17SW, supporting the idea that they were formed through the interaction between the H\,{\sc ii} region and the molecular cloud. 
Recent studies reported evidence for still accreting massive or intermediate-mass protostars for a couple of objects which were identified as silhouette disks sometimes associated with H$_2$ jets \citep[e.g.][]{Nuernberger,Niel07,Niel08}.

The distance of the M\,17 complex remains somewhat uncertain. From multi-color photometry and spectroscopy, \citet{Hoff} inferred a distance of $2.1 \pm 0.2$\,kpc and a ratio of total to selective extinction of $R_V = 3.9$. These authors argued that previous photometric and spectrophotometric distance estimates, around 1.6\,kpc, were probably biased by ignoring the multiplicity of O-type stars in NGC\,6618 \citep[see also the discussion in][]{Povich09}. 

The total extinction towards M\,17 consists of a foreground extinction, a local dust component inside M\,17 and, in some cases, absorption by circumstellar disks or dust. This results in a very large optical extinction with a very complex spatial distribution \citep{Povich09}. 

The stellar population of M\,17 features a number of very young, hot and massive objects. The main exciting source of M\,17 is \astrobj{CEN\,1} \citep{CEN}, also known as {\it Kleinmann's Anonymous Star}\footnote{Throughout this paper, we adopt the naming convention of \citet[][CEN designations]{CEN} and \citet[][OI designations]{OI}.}. CEN\,1 is a Trapezium-like system consisting of two spectroscopic binaries (CEN\,1a and 1b) that form a visual pair with an angular separation of 1.8\,arcsec. Each of these spectroscopic binaries is of spectral type O4\,V \citep{Hoff}\footnote{The physical association of the two binaries is still controversial since they are subject to different optical extinctions \citep{Hoff}.}. Actually, multiplicity might be quite common among the O-type stars of NGC\,6618. Indeed, \astrobj{CEN\,18} (O6-8\,V) was also found to display double absorption lines, and \astrobj{CEN\,37} (O3-6\,V), being over-luminous for its spectral type, was suggested to be another binary candidate \citep{Hoff}. VLA radio observations at 3.6 and 6\,cm wavelength revealed several compact radio sources, two of them being associated with the components of the visual binary CEN\,1 \citep{RGM}. CEN\,1a is detected only at the shortest wavelength. Both radio sources, especially the one associated with CEN\,1a, have flux densities that exceed the expected level of free-free emission from their stellar winds. \citet{RGM} accordingly suggest that the observed radio emission could, at least partially, stem from the synchrotron process in a colliding wind region of the binary system.

Although stellar winds account only for a very small fraction (a few percent at most) of the overall energy input of early-type stars into H\,{\sc ii} regions, their mechanical power determines the dynamics of a nebula and produces so-called wind-blown bubbles or superbubbles (in the case of clusters of early-type stars). The superbubble blown by the stellar winds of the OB stars in M\,17 is highly asymmetric due to its interaction with the neighboring molecular cloud M\,17SW. Soft X-ray emission spatially coincident with the superbubble of M\,17 was detected with previous X-ray observations using {\it ROSAT} \citep{Dunne}, {\it Chandra} \citep{Townsley1, Townsley2} and {\it Suzaku} \citep{Hyodo}. Beside this diffuse emission, these X-ray observations also revealed a number of point sources. For instance, a 40\,ksec {\it Chandra}-ACIS observation led to the detection of 886 point-like sources in the field of M\,17 \citep{Broos}. 

In the present paper we discuss the results of an {\it XMM-Newton} observation of M\,17 with the goal to quantify the properties of the brightest X-ray sources in the field. Section\,\ref{obs} provides an overview of our observations. In Sect.\,\ref{sources}, we analyze the X-ray sources found in the field of M\,17. We cross-correlate their position with various catalogs to identify counterparts at other wavelengths. We then analyze the spectra of twenty X-ray bright point like sources as well as of the diffuse X-ray emission. Finally, in Sect.\,\ref{conclusions}, we summarize the most important results of our study and present our conclusions.  
\begin{figure}[h!]
\begin{center}
\resizebox{8cm}{!}{\includegraphics{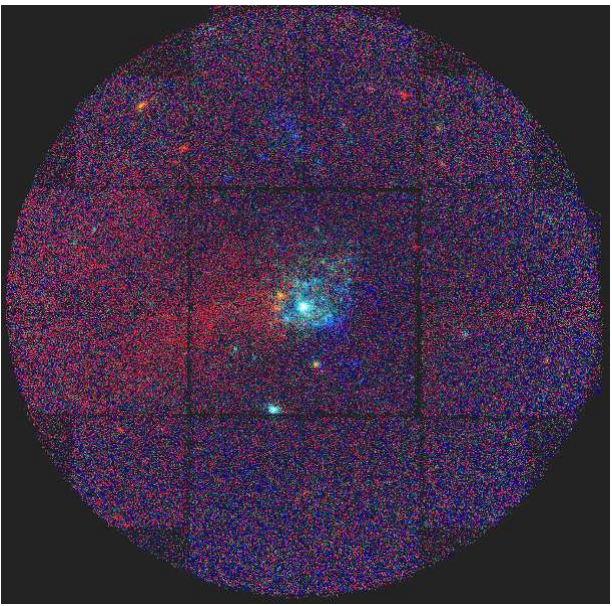}}
\end{center}
\caption{Energy-coded three-color X-ray image of M\,17 built from the EPIC-MOS data. North is up and east is to the left. The field of view of the EPIC instruments is centered on CEN\,1 and has a radius of 15\,arcmin. The red, green, and blue colors correspond to the soft, medium, and hard energy bands used throughout this paper (see Sect.\,\ref{obs}). The individual images were exposure corrected, before they were combined. Note the reddish emission lane that extends to the east-south-east of the central cluster. \label{X3col}}
\end{figure}
\begin{figure}[h]
\vspace*{3mm}
\centering
\includegraphics[width=80mm]{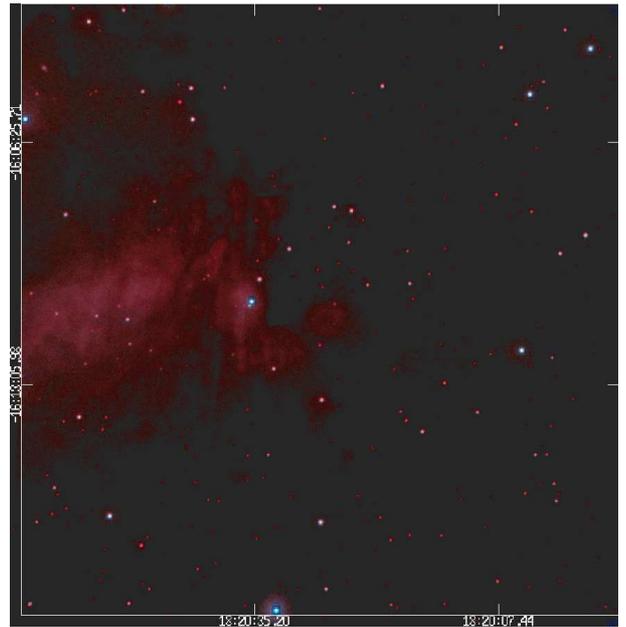}
\caption{Energy-coded three color UV image of M\,17 as built from the $UVW1$ (red), $UVM2$ (green) and $UVW2$ (blue) OM data. The OM field of view covers a region of $17 \times 17$\,arcmin$^2$, centered on CEN\,1. Note the diffuse emission that is seen eastward of the center. The brightest point-like sources are affected by some artifacts (see Sect.\,\ref{obs}).}
\label{OM3col}
\end{figure}

\section{Observations \label{obs}}
M\,17 was observed for 30\,ksec with the {\it XMM-Newton} satellite \citep{Jansen} on 3 November 2003 (JD\,2\,452\,710.170 -- 2\,452\,710.520). The two EPIC-MOS detectors \citep{MOS} and the EPIC-pn camera \citep{pn} were all used in full frame mode with the medium filter to reject optical light. 

The raw data were processed with the SAS software (version 10.0). Our observation was affected by a high radiation background. As the background showed no isolated flares, but rather a general high level affecting the entire duration of the observation, we did not define good time intervals and kept the entire observation instead. 

Images were extracted for soft (0.5 -- 1.0\,keV), medium (1.0 -- 2.0\,keV), and hard (2.0 -- 10.0\,keV) energy bands. An energy-coded three-color image of the combined field of view of the two EPIC-MOS instruments is shown in Fig.\ \ref{X3col}. In this image, we have emphasized the soft band to highlight the weak diffuse emission that extends to the east-south-east of the central cluster.

\begin{figure*}
\centering
\begin{minipage}{9cm}
\includegraphics[width=8cm]{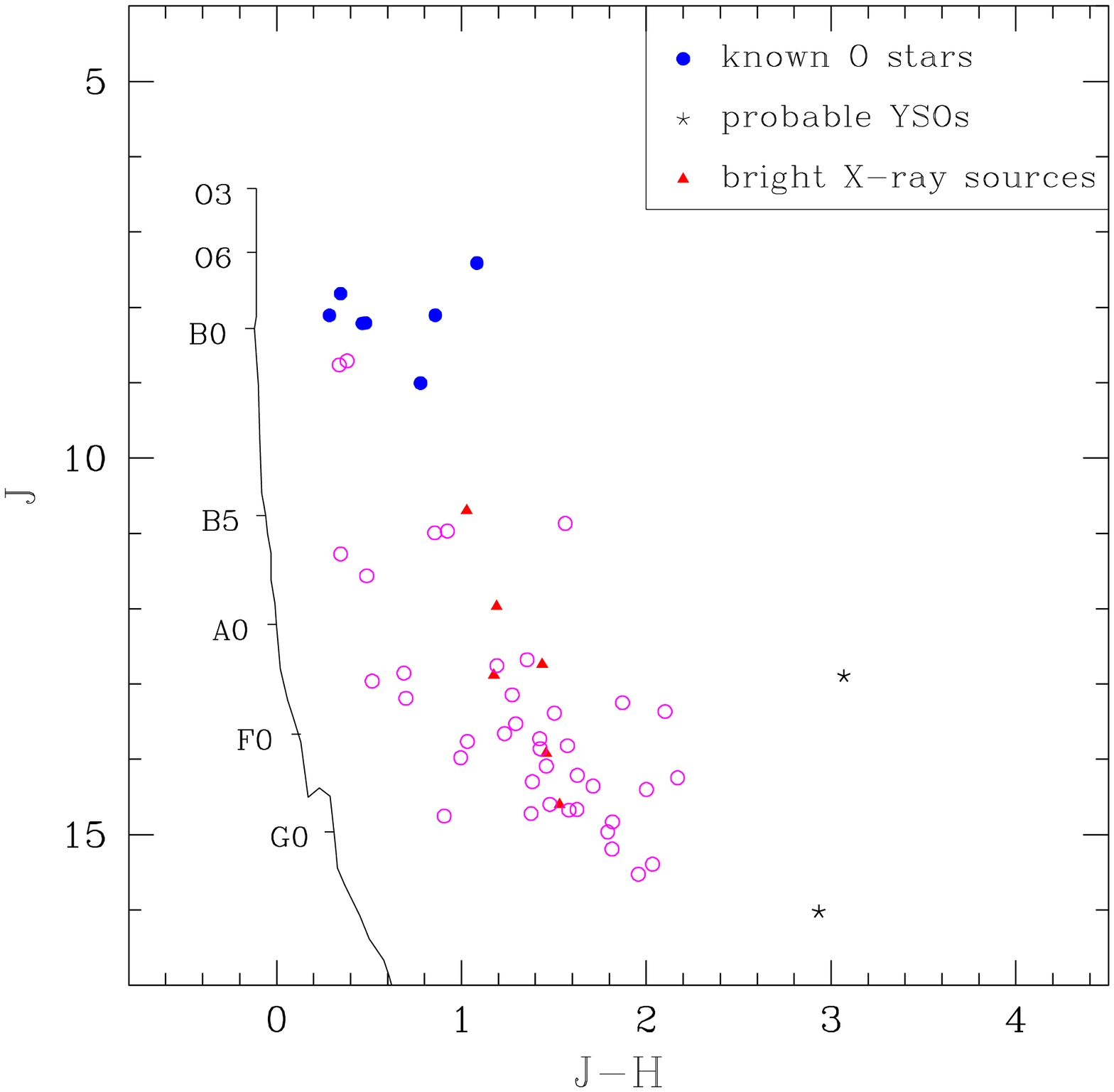}
\end{minipage}
\hfill
\begin{minipage}{9cm}
\includegraphics[width=8cm]{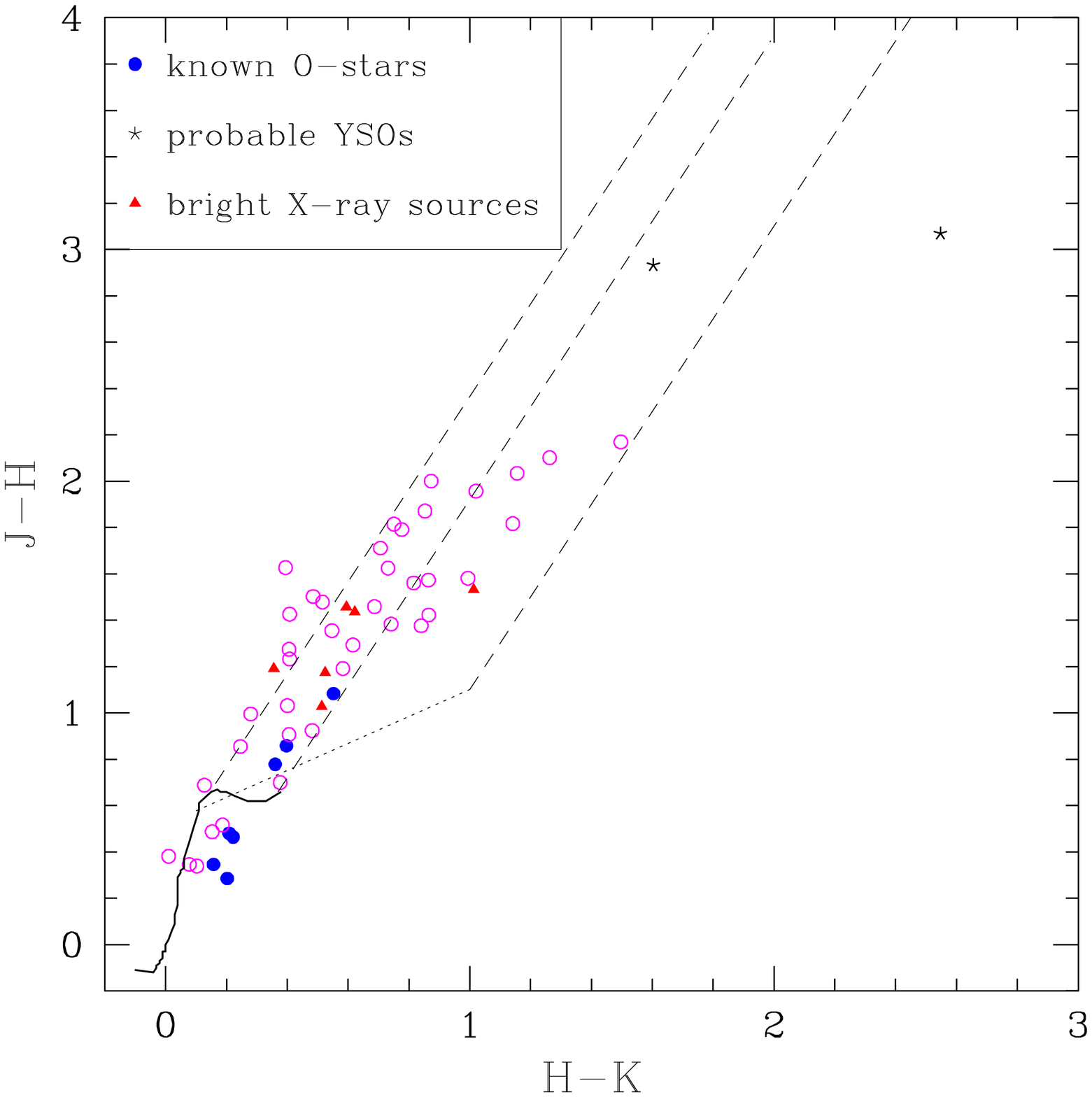}
\end{minipage}
\caption{Color-magnitude (left) and color-color (right) diagrams of the near-IR counterparts of the X-ray sources detected with EPIC in the M\,17 field of view. Only 2MASS data with reliable photometry are plotted. The known O-stars and the bright X-ray sources are objects that are studied spectroscopically in Sect.\,\ref{spectroX}. The slope of the reddening relation in the color-color diagram, indicated by the dashed lines, is taken to be $E(J - H)/E(H - K) = 2$ in agreement with \citet{Hoff} and \citet{SL}. The solid line yields the locus of the un-reddened main-sequence according to \citet{Tokunaga} and \citet{MartinsPlez} and shifted by a distance modulus of 11.61 in the color-magnitude diagram. The dotted straight line in the color-color diagram indicates the locus of the un-reddened classical T-Tauri stars following \citet{Meyer}.}
\label{2MASS}
\end{figure*}

The Optical Monitor \citep[OM,][]{OM} instrument aboard {\it XMM-Newton} observed M\,17 through the three UV filters $UVW1$, $UVM2$ and $UVW2$ with exposure times of 4, 4 and 8\,ksec respectively. The OM was operated in `full frame low-resolution' mode covering a $17 \times 17$\,arcmin$^2$ square field of view centered on the same coordinates as the X-ray cameras. The $UVW1$, $UVM2$ and $UVW2$ filters have bandwidths of 2450 -- 3200, 2050 -- 2450, and 1800 -- 2250\,\AA, respectively \citep{OM}.

The OM data were processed with the relevant command chain of the SAS software. The data from the three filters were used to build a color-coded image (Fig.\ \ref{OM3col}). The OM images exhibit some artifacts such as smoke-ring ghosts caused by light reflected off a chamfer in the detector window and modulo-8 pattern around bright sources \citep{OM}. Nevertheless, Fig.\,\ref{OM3col} reveals also some genuine diffuse emission in the $UVW1$ filter. A total of 614 point sources are detected in the OM images (a Table with all the detections will be made available electronically), 128 of them being detected in all three passbands. 

The OM detector has a read-out time of 11\,ms and sources with a count rate of 10\,counts\,s$^{-1}$ typically suffer from coincidence losses of about 10\% \citep{OM}. These losses are partially recovered during the data processing to provide `corrected' count rates, which are then converted into AB magnitudes \citep{Oke}. With the filters used during our observation, corrected count rates of 10\,counts\,s$^{-1}$ correspond to $UVW1$, $UVM2$ and $UVW2$ AB-magnitudes of 16.066, 14.912 and 14.072, respectively. The reconstructed photometry of sources much brighter than this should be considered with caution.

\section{X-ray sources in M\,17 \label{sources}}
Figure\,\ref{X3col} shows that the X-ray emission of M\,17 is dominated by a few bright point sources along with a number of fainter point sources and some diffuse emission. The SAS detection routines produced a list of 99 sources in the combined EPIC images with a significance threshold\footnote{This implies a probability of $\leq e^{-10}$ that a random Poissonian fluctuation could have caused the observed source counts inside the detection cell.} of 10. 

All these sources were inspected by eye and 16 of them were rejected as spurious. This inspection revealed another 9 sources that were clearly seen on the image, but were not found by the detection algorithm. The reasons the detection routines did a rather poor job on these data are essentially the diffuse background as well as the generally high level of particle background. We ended up with a final list of 92 sources (a table with these detections will be made available electronically). This is far less than the number of sources (886) detected by \citet{Broos} with {\it Chandra}. The main reasons for this situation are the wider point spread function of {\it XMM-Newton} compared to {\it Chandra} and the high background level of our observation. From a histogram of the number of sources versus count rate, we conclude that our detections are complete down to rates of 0.004 and 0.002\,cts\,s$^{-1}$ for the EPIC-pn and the individual EPIC-MOS cameras, respectively.

\begin{table*}
\caption{Results of the cross-correlation between the list of X-ray sources detected with our EPIC data of M\,17 and the list of OB stars from \citet{Povich09}. Spectral types with a colon are uncertain, and could actually be earlier than quoted here.\label{PovichX}}
\begin{center}
\begin{tabular}{c c c l c c c c c}
\hline
XID & $\alpha$ (J\,2000.0) & $\delta$ (J\,2000.0) & Name & Spectral type & A$_V$ & $UVW2$ & $UVM2$ & $UVW1$\\
\# & hh:mm:ss & deg:arcmin:arcsec & & & mag \\
\hline
38 & 18:20:25.9 & $-$16:08:33 & \astrobj{CEN\,18} & O6-8\,V & 7.6 & & & $18.85 \pm 0.05$ \\
45 & 18:20:27.5 & $-$16:13:32 & \astrobj{OI\,345} & O6\,V   &      & $16.12 \pm 0.02$ & $16.05 \pm 0.01$ & $13.94 \pm 0.01$ \\
55 & 18:20:29.8 & $-$16:10:45 & \astrobj{CEN\,1a,b} & O4\,V + O4\,V & &  & & $19.73 \pm 0.11$ \\
68 & 18:20:34.5 & $-$16:10:12 & \astrobj{CEN\,2}  & O5\,V & 5.2  & $15.95 \pm 0.01$ & $15.83 \pm 0.01$ & $13.77 \pm 0.01$ \\
71 & 18:20:35.4 & $-$16:10:49 & \astrobj{CEN\,3}  & O9\,V: & 3.7 & $13.59 \pm 0.01$ & $13.51 \pm 0.01$ & $13.05 \pm 0.01$ \\
72 & 18:20:35.9 & $-$16:15:43 & \astrobj{OI\,352} & O8\,V: & 7.0 & $20.17 \pm 0.23$ & $20.19 \pm 0.19$ & $17.65 \pm 0.02$ \\
88 & 18:21:02.2 & $-$16:01:01 & \astrobj{BD$-16^{\circ}$\,4826} & O5 & 3.9 \\
\hline
\end{tabular}
\end{center}
\end{table*}
\normalsize

\subsection{Cross-correlation with other catalogs}
To start, we have cross-correlated the list of X-ray sources with the 2MASS point sources catalog \citep{2mass,Skrutskie}. Using the technique of \citet{Jeff97}, the optimal cross-correlation radius, which offers the best compromise between obtaining the maximum number of true identifications and avoiding contamination by spurious coincidences, was established to be 2.5\,arcsec. For this correlation radius, we expect to find fewer than 7\% of spurious cross-identifications among the 67 positive matches listed. 

The $J$ versus $J-H$ color-magnitude and $J - H$ versus $H - K$ color-color diagrams of the 2MASS counterparts of the sources detected with EPIC are shown in Fig.\,\ref{2MASS}. The main-sequence relation in these figures is taken from \citet{Tokunaga} for stars later than spectral type O, and from \citet{MartinsPlez} for O-type stars. As can be seen on these diagrams, there is a wide range of reddening values, which hampers a direct comparison with theoretical isochrones. We note that about 1/6 of the X-ray sources have a near-IR counterpart that falls in the region occupied by classical T-Tauri stars (cTTs) which exhibit IR-excess emission from a circumstellar disk \citep{Meyer}. This result is in excellent agreement with the finding of \citet{Broos} who noted that whilst most of the {\it Chandra} sources occupied the near-IR color space associated with reddened weak-line T\,Tauri stars, about one sixth of their sources fell in a region of the near-IR color-color diagram where classical T\,Tauri and Herbig AeBe stars are expected. This contrasts with the situation in some other young open clusters (e.g.\ Cyg\,OB2, \citeauthor{Rauw}\,\citeyear{Rauw}; NGC\,6231, \citeauthor{Sana}\,\citeyear{Sana}; and NGC\,6383, \citeauthor{NGC6383}\,\citeyear{NGC6383}), where only a very small fraction of the counterparts of faint X-ray sources provide evidence for near-IR excesses.\\

We have cross-correlated our list of sources with the list of OB stars provided by \citet{Povich09}. 16 known O-type stars fall within the field of view covered by the EPIC instruments. Within a correlation radius of 2.5\,arcsec, we found 7 positive correlations (see Table\,\ref{PovichX}). Except for \astrobj{CEN\,3} and \astrobj{OI\,352}, all of them have a spectral type earlier than O7. The non-detections mostly refer to stars later than O7. There are a few exceptions though. CEN\,37 (O3-6\,V) is located very close to the pair CEN\,1a,b and could contribute to the X-ray emission associated with the latter, without being resolved as an individual source. CEN\,43 (O3-5\,V) falls close (6.9\,arcsec) to one of the brightest X-ray sources (\# 64) in our observation, but their association is formally not significant. Whilst we cannot rule out that CEN\,43 slightly contributes to the emission of source \#64, it must be stressed though that this star is very heavily reddened \citep[$A_V = 12.3$,][]{Povich09} and could thus escape detection simply as a result of the interstellar absorption. 

In addition, this cross-correlation identifies the two most heavily reddened sources in Fig.\,\ref{2MASS} as being two candidate young stellar objects (YSOs) previously seen with {\it Chandra}: our X-ray sources 17 and 75 being respectively associated with objects X4 and X14 in \citet{Povich09}.\\

Correlation of our X-ray sources with the catalog of OM point sources yielded a rather low number of matches: only 11 of the 76 EPIC sources that fall into the OM field of view were found to have an OM counterpart. Six of these objects are OB stars (see Table\,\ref{PovichX}). The poor match between the X-ray sources and the UV detections is yet another manifestation of the huge absorption towards M\,17.

\begin{figure*}
\centering
\begin{minipage}{6cm}
\includegraphics[width=6cm]{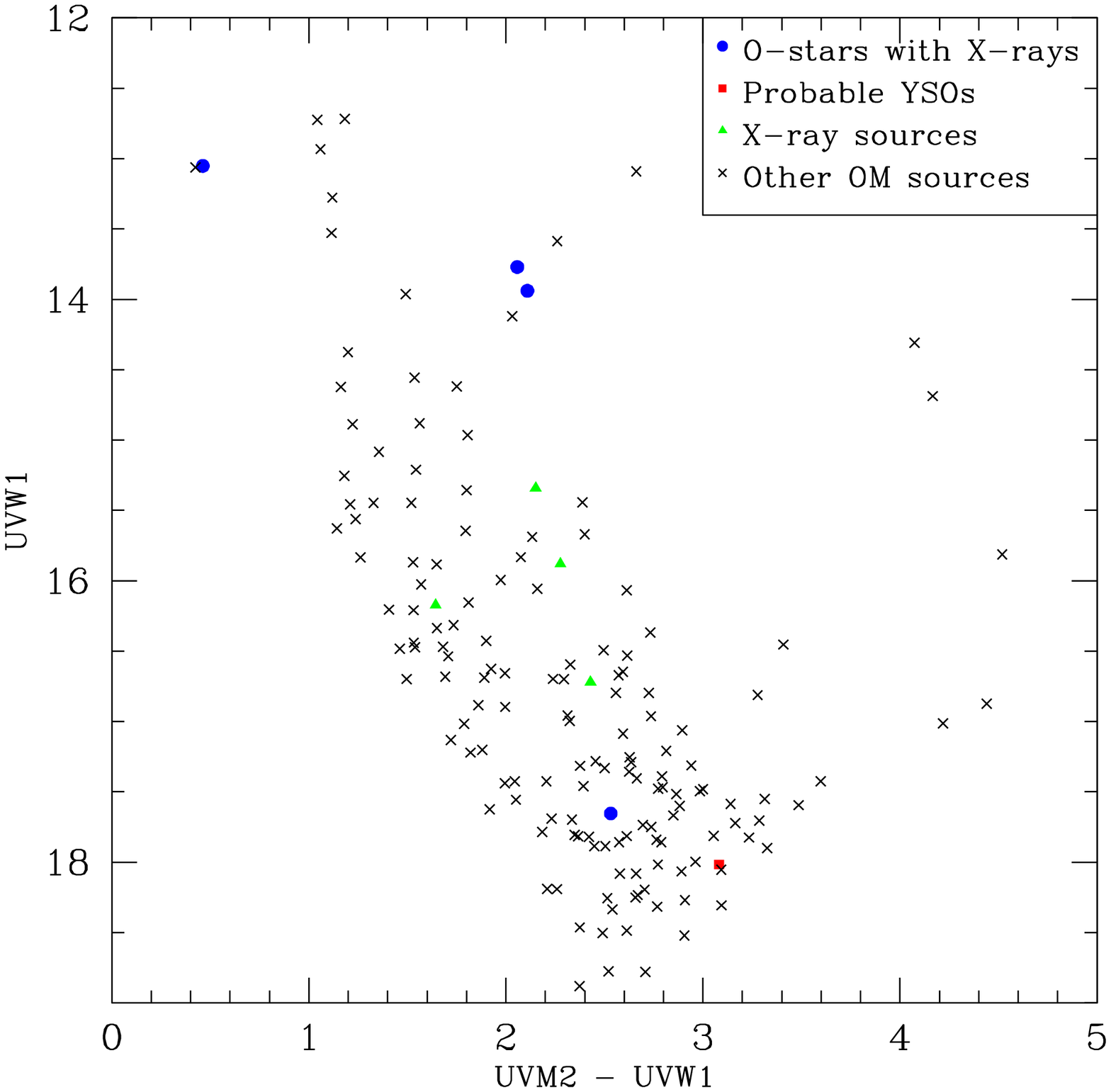}
\end{minipage}
\hfill
\begin{minipage}{6cm}
\includegraphics[width=6cm]{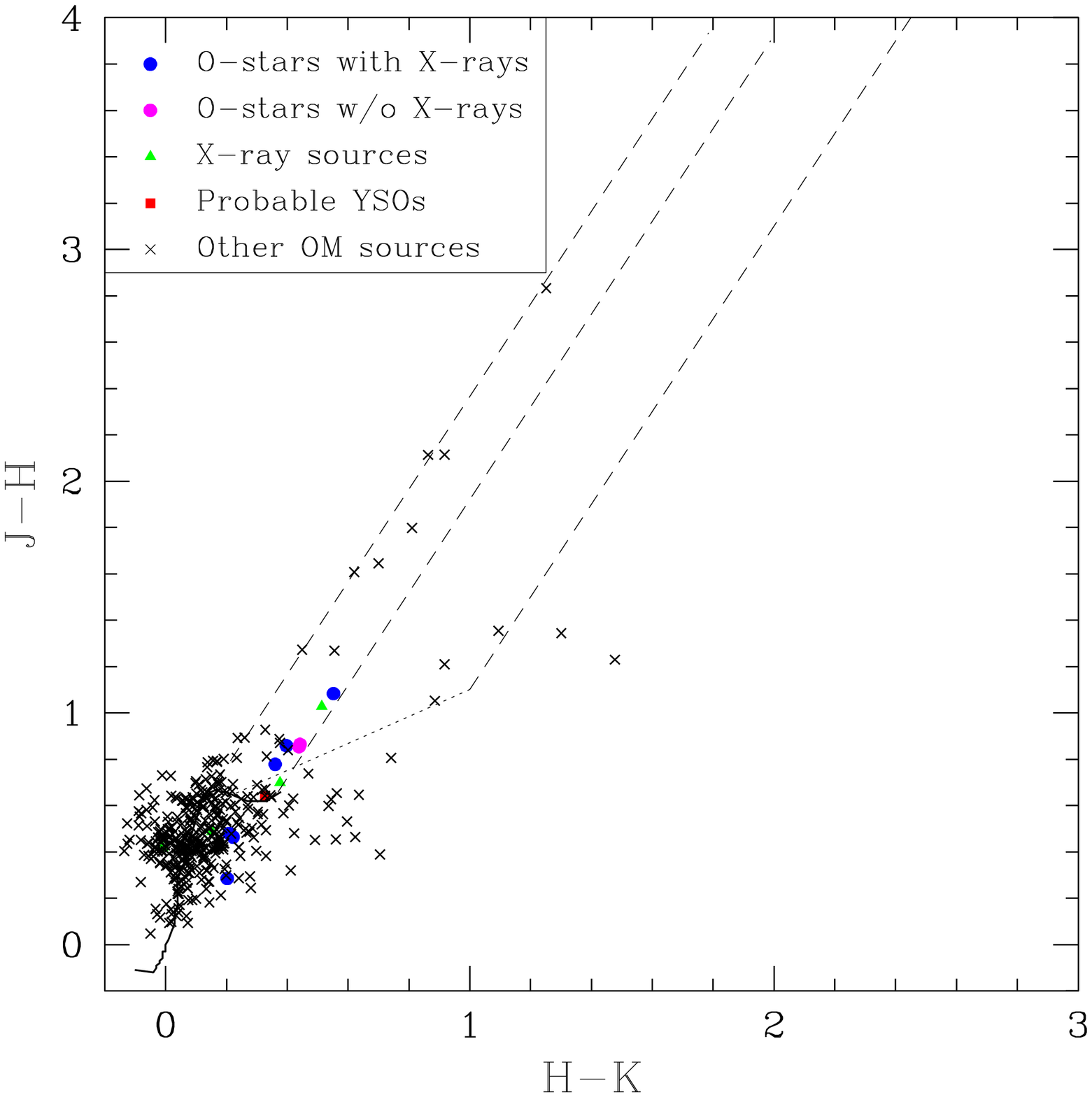}
\end{minipage}
\hfill
\begin{minipage}{6cm}
\includegraphics[width=6cm]{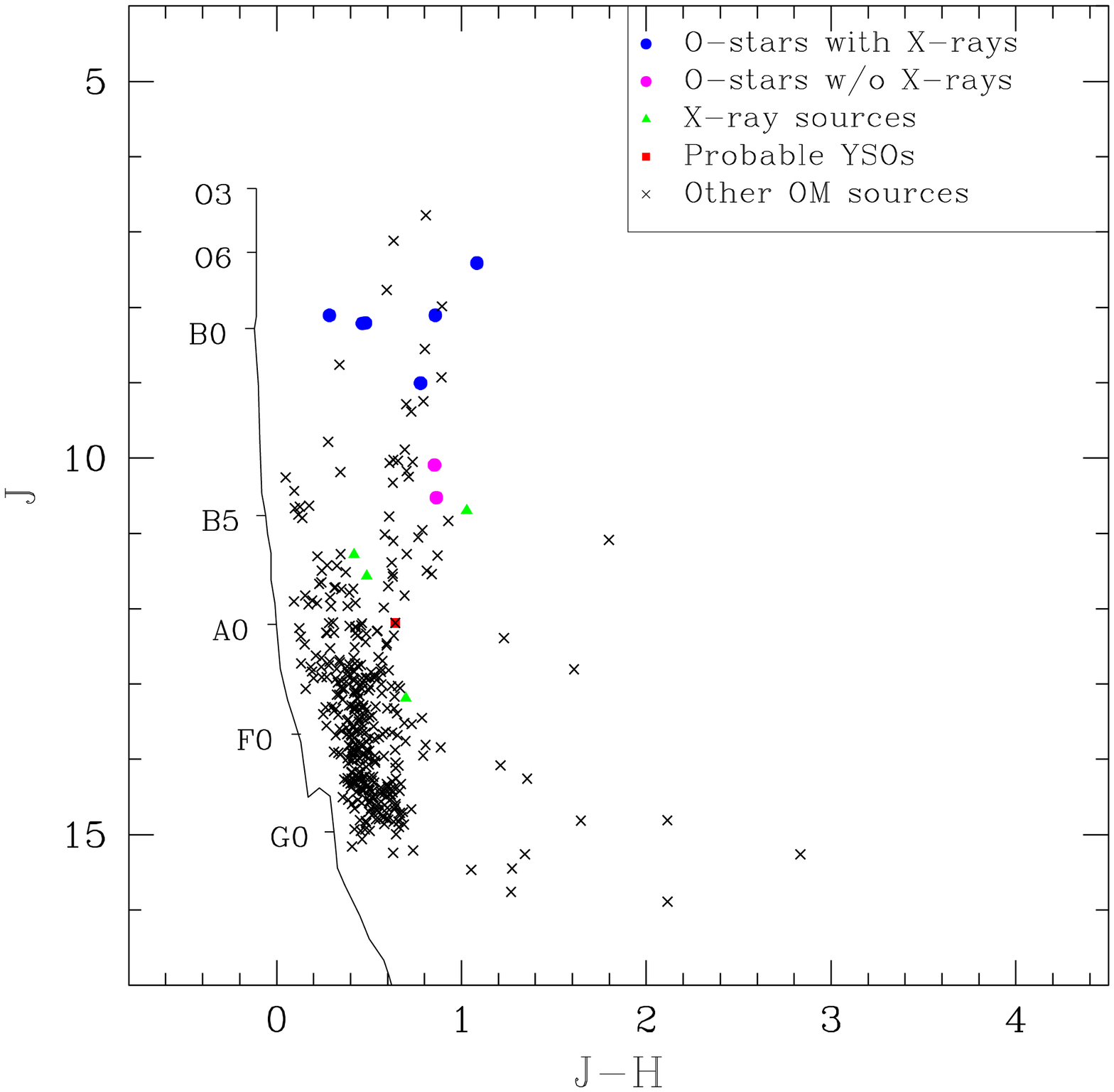}
\end{minipage}
\caption{Left: UV color-magnitude diagram of the OM sources detected in the $UVW1$ and $UVM2$ filters. Middle and right: color-color and color-magnitude diagrams of the near-IR counterparts of the UV sources detected with the OM in the M\,17 field of view. Only 2MASS data with reliable photometry are plotted. The slope of the reddening relation and the main-sequence relations are built in the same way as in Fig.\,\ref{2MASS}.}
\label{OM2MASS}
\end{figure*}

This brings up the question about the nature of the OM sources. Cross-correlation of the OM sources with the 2MASS point source catalog yields an optimal correlation radius in the range 2 -- 2.5\,arcsec. Adopting a radius of 2.5\,arcsec, 510 of the 614 OM point sources have a near-IR counterpart. The corresponding $J-H$ versus $H -K$ color-color diagram as well as the near-IR and UV color-magnitude diagram (see Fig.\,\ref{OM2MASS}) clearly indicate that the majority of these objects are less-reddened, likely foreground late-type stars, unrelated to M\,17.

Comparing the list of OM sources with the catalogs of OB stars and candidate young stellar objects of \citet{Povich09} yields 8 positive matches for OB stars and one YSO within a radius of 2.5\,arcsec. Two of the OB stars seen with the OM are not detected in X-rays. These are CEN\,25 and 31 of spectral type O7-8\,V and O9.5\,V respectively.\\

\begin{figure}
\centering
\includegraphics[width=9cm]{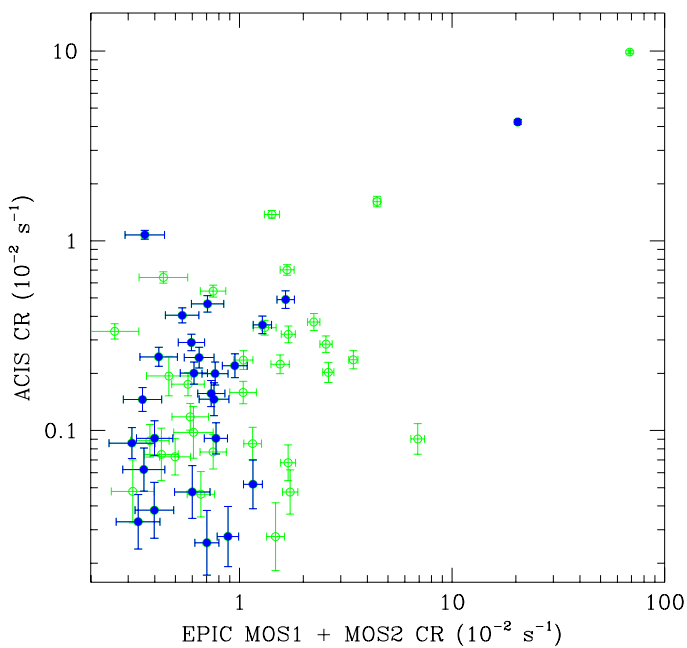}
\caption{Comparison of the EPIC-MOS and ACIS-I count rates for the X-ray sources in common between \citet{Broos} and the present paper. Open and filled symbols stand for EPIC sources which have a single ACIS counterpart within a correlation radius of respectively 2.5 and 10\,arcsec. }
\label{ChandravsEPIC}
\end{figure}

Finally, we have cross-correlated the list of our X-ray sources with the X-ray sources detected by the {\it Chandra}-ACIS observation \citep{Broos}. Adopting a correlation radius of 2.5\,arcsec, we find 61 positive correlations. Actually, some of our brightest EPIC sources do not have an ACIS counterpart (within the correlation radius) and that some of the prominent ACIS sources are missing in our list of detections. We have then used these results to compare the ACIS and EPIC count rates\footnote{For the latter we adopted the MOS instruments as they have fewer gaps between CCDs than the pn detector, thus allowing a reliable determination of the count rates for a larger number of sources.}. Figure\,\ref{ChandravsEPIC} reveals a highly scattered relation, that deviates significantly from a simple proportionality. One can consider several explanations for the large scatter:
\begin{itemize}
\item[$\bullet$] there could be a bias in the EPIC count rates due to source confusion. Indeed, the density of X-ray sources as seen with {\it Chandra} is quite high, especially in the crowded core of the NGC\,6618 cluster. Due to the coarser PSF of {\it XMM-Newton} (as compared to {\it Chandra}) several {\it Chandra} sources could be confused and counted as a single source by {\it XMM-Newton}. To test this scenario, we have repeated the cross-correlation with a radius of 10\,arcsec and we have replotted only those EPIC sources with a single ACIS counterpart within 10\,arcsec (see Fig.\,\ref{ChandravsEPIC}). This treatment should eliminate the bulk of the sources that are due to confusion, but we see little effect on the distribution of the objects in Fig.\,\ref{ChandravsEPIC}. Therefore, confusion alone cannot account for the scatter in the count rate relation.
\item[$\bullet$] the brightest sources could suffer from pile-up. Because of {\it Chandra}'s very narrow PSF, this problem would mainly impact the ACIS-I count rates. However, for all but the two brightest sources in Fig.\,\ref{ChandravsEPIC}, pile-up should be of a few percent at most. Indeed, for the brightest and second brightest objects in this figure, we estimate pile-up levels of about 12\%, and 5\% respectively. For sources with less than 0.01\,count\,s$^{-1}$, the pile-up level should be less than 1.2\%.
\item[$\bullet$] there could be systematics due to the instruments. Whilst one cannot rule out some residual problems in the cross-calibration of the EPIC and ACIS-I instruments, it has to be noted that a major effort has been put into this task over the last decade. All major issues have been solved and remaining problems are mostly small. Of course, some of the scatter in the count rates relation could reflect the different sensitivities of the EPIC-MOS and ACIS-I detectors. However, this should be a rather small effect.
\item[$\bullet$] there could be intrinsic time variability of the faint sources on long time scales. The {\it Chandra} and {\it XMM-Newton} observations were taken twenty months apart. It seems quite plausible that the X-ray brightness of weak sources, which could be associated with low-mass pre-main sequence stars, changes with time. In view of Fig.\,\ref{ChandravsEPIC} and the above discussion, this appears to be most likely explanation of the scatter and deviation from a simple proportionality. A spectacular illustration of this scenario is provided by source 396 of \citet{Broos}. This is the second brightest X-ray source in the {\it Chandra} observation, but is not detected in our observation\footnote{The closest EPIC source to this object is our source \#42 at a separation of 4.9\,arcsec.}. Conversely, our fourth brightest source (XID \# 64) has no equivalent in the ACIS-I data, the nearest possible ACIS counterparts being very faint and located at more than 4.5\,arcsec. 
\end{itemize}

\subsection{Spectral analyses \label{spectroX}}
The EPIC spectra of twenty point-like sources were extracted and fitted using the {\tt xspec} software \citep[version 12.6.0,][]{Arnaud}. These include all the O-stars detected with EPIC as well as the brightest objects (with EPIC-MOS count rates of at least 0.01 count\,s$^{-1}$) in the field of view. The source regions were chosen in such a way to include the majority of the source photons, whilst simultaneously avoiding overlap with nearby sources in the most crowded regions of the field. The background spectra were evaluated from nearby circular regions, free of any obvious point source.  

For all sources, we tested absorbed power-law models as well as absorbed thermal plasma {\tt apec} models \citep{apec}. For the latter, we first assumed solar abundances. However, for the best exposed spectra, including the bright CEN\,1 system, this resulted in a rather poor fit especially of the Fe\,K line. We then let the global metallicity vary during the fit and, in most cases, it converged to a value 0.2 -- 0.4 times solar \citep{Mernier}. However, given the limited quality of most X-ray spectra, a more robust approach would be to constrain the abundances independently and then fit the X-ray spectra using fixed abundances. 

\citet{Garcia} studied the chemical composition of the M\,17 H\,{\sc ii} region and determined abundances of nine key elements (including He, C, N, O, Ne, and Fe). Compared to the solar composition \citep{asplund09}, most elements are found to have abundances $\sim 0.7$ times solar. There are a few notable exceptions to this rule. These are He, which has a solar abundance, C, which is found to be 2.4 times solar and, most of all, Fe, which is found to have an abundance of 0.026 solar. \citet{Garcia} show that their results are generally in good agreement with those from the literature, including for the iron abundance. The abundances of He and C have no directly measurable impact on the EPIC spectra of our sources: helium has no spectral features inside the EPIC energy domain, whilst carbon has some lines that fall at the low energy edge of the EPIC sensitivity range, but are of little relevance here, because of the huge interstellar absorption towards M\,17. However, iron is of major importance for the EPIC spectra. Using such a low Fe abundance in our spectral models actually results in extremely poor fits. After some trials, we thus decided to fix all metal abundances to 0.7 times solar, except for He, C and Fe, which were fixed at 1.0, 2.4 and 0.4 times solar respectively. 

\subsubsection{X-ray spectra of O-type stars \label{XspecO}} 
For the O-type stars listed in Table\,\ref{PovichX}, we have constrained the value of the interstellar neutral hydrogen column density using the relation between $N_H$ and $E(B-V)$ derived by \citet{BSD}. For this purpose, we adopt the values of $A_V$ quoted in Table\,\ref{PovichX} and taken from \citet{Povich09}\footnote{We note that these $A_V$ values are larger than those quoted by \citet{Hoff} for the stars in common among our sample.}, accounting for the peculiar reddening law $R_V = 3.9$ \citep{Hoff}. 
For CEN\,1a and CEN\,1b, there are two different values of $A_V$ of respectively 10.2 and 13.5\,mag \citep{Hoff}. In our fits of the X-ray spectra of O-stars, we allow for some excess absorption due to the ionized stellar wind \citep{HD108} or to some circumstellar absorption. Therefore, we choose the lower of the two $A_V$ values for constraining the $N_H$ of the spectrum of source \# 55 (corresponding to CEN\,1a + 1b). There is no estimate of $A_V$ for OI\,345 in the catalog of \citet{Povich09}. However, the source is detected in all three OM filters and by comparison with CEN\,2 (which has a similar spectral type), we estimate an extinction of $A_V \simeq 5.3$ for OI\,345\footnote{We note that \citet{Hoff} quote $A_V = 4.3$ and $4.0$ respectively for OI\,345 and CEN\,2.}.   

\begin{table*}
\caption{Results of the fits of the EPIC spectra of the O-type stars of M\,17 with models of the kind {\tt wabs*wind*vapec} and {\tt wabs*wind*vapec(2T)}. The interstellar neutral hydrogen column densities were frozen at the values given in column 2. The absorption by the ionized wind is accounted for using the model of \citet{HD108}. The abundances of the hot plasma were taken to be 1.0, 2.4, 0.4 and 0.7 times solar for He, C, Fe and all other metals, respectively (see text). A two temperature plasma model is only required for CEN\,1a,b and OI\,352. In these cases, column six quotes the ratio of the emission measures of the hotter and cooler components. Column 7 provides the reduced $\chi^2$ of the fit as well as the number of degrees of freedom (in brackets). The observed and absorption-corrected X-ray fluxes are evaluated in the 0.5 -- 10\,keV energy range.\label{fitsO}}
\begin{center}
\begin{tabular}{l c c c c c c c c}
\hline
Star & $N_H$ & $\log{N_{\rm wind}}$ & kT$_1$ & kT$_2$ & EM$_2$/EM$_1$ & $\chi^2_{\nu}$ (d.o.f.) & $f_X$ & $f_X^{\rm unabs}$ \\
     & $10^{22}$\,cm$^{-2}$ & cm$^{-2}$ & keV & keV &           &      & $10^{-13}$\,erg\,cm$^{-2}$\,s$^{-1}$ & $10^{-13}$\,erg\,cm$^{-2}$\,s$^{-1}$ \\
\hline
CEN\,18 & 1.13 & $22.1^{+0.9}_{-1.9}$ & $0.3^{+.4}_{-.2}$ & -- & -- & 1.13 (27) & 0.11 & 1.25 \\
OI\,345 & 0.79 & $\leq 21.6$ & $0.6^{+.1}_{-.1}$ & -- & -- & 0.91 (29) & 0.51 & 3.22 \\
CEN\,1a,b & 1.52 & $21.3^{+0.1}_{-0.2}$ & $0.61^{+.04}_{-.03}$ & $3.8^{+.2}_{-.2}$ & $0.70 \pm 0.06$  & 1.13 (455) & 42.30 & 129.60 \\
CEN\,2 & 0.77 & $21.6^{+0.2}_{-0.3}$ & $0.30^{+.05}_{-.06}$ & -- & -- & 0.91 (74) & 0.81 & 9.28 \\
CEN\,3 & 0.55 & $21.5^{+0.3}_{-0.9}$ & $0.30^{+.09}_{-.06}$ & -- & -- & 1.43 (49) & 0.28 & 2.09 \\
OI\,352 & 1.04 & $21.70^{+0.2}_{-0.2}$ & $0.6^{+.4}_{-.2}$ & $2.6^{+.3}_{-.3}$ & $1.15 \pm 0.27$ & 1.05 (234) & 11.06 & 22.87 \\
BD$-16^{\circ}$\,4826 & 0.58 & $21.7^{+0.4}_{-0.3}$ & $0.31^{+.09}_{-.08}$ & -- & -- & 1.21 (29) & 0.64 & 4.28 \\
\hline
\end{tabular}
\end{center}
\end{table*}
\normalsize

\begin{figure*}
\centering
\begin{minipage}{9cm}
\includegraphics[width=9cm]{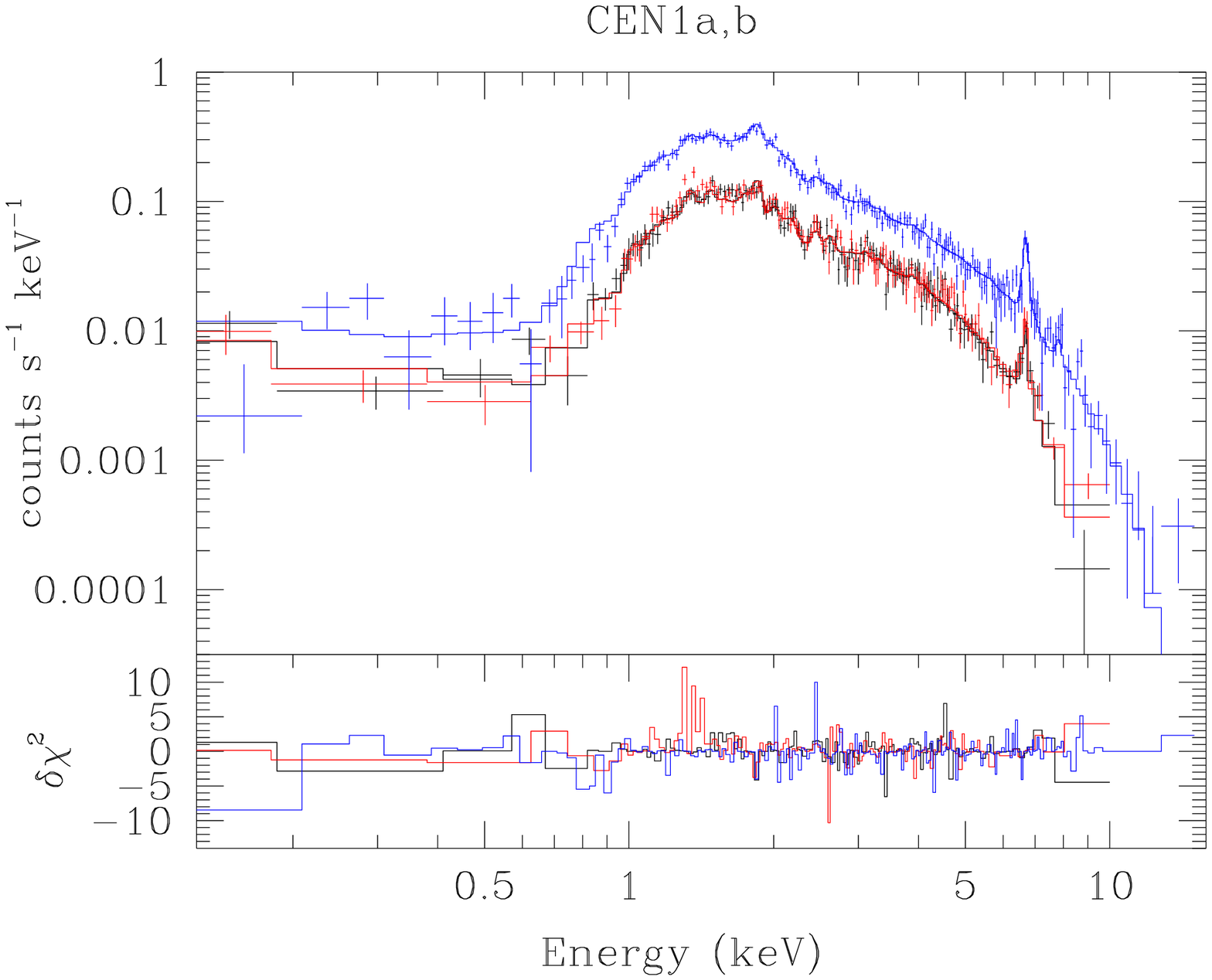}
\end{minipage}
\hfill
\begin{minipage}{9cm}
\includegraphics[width=9cm]{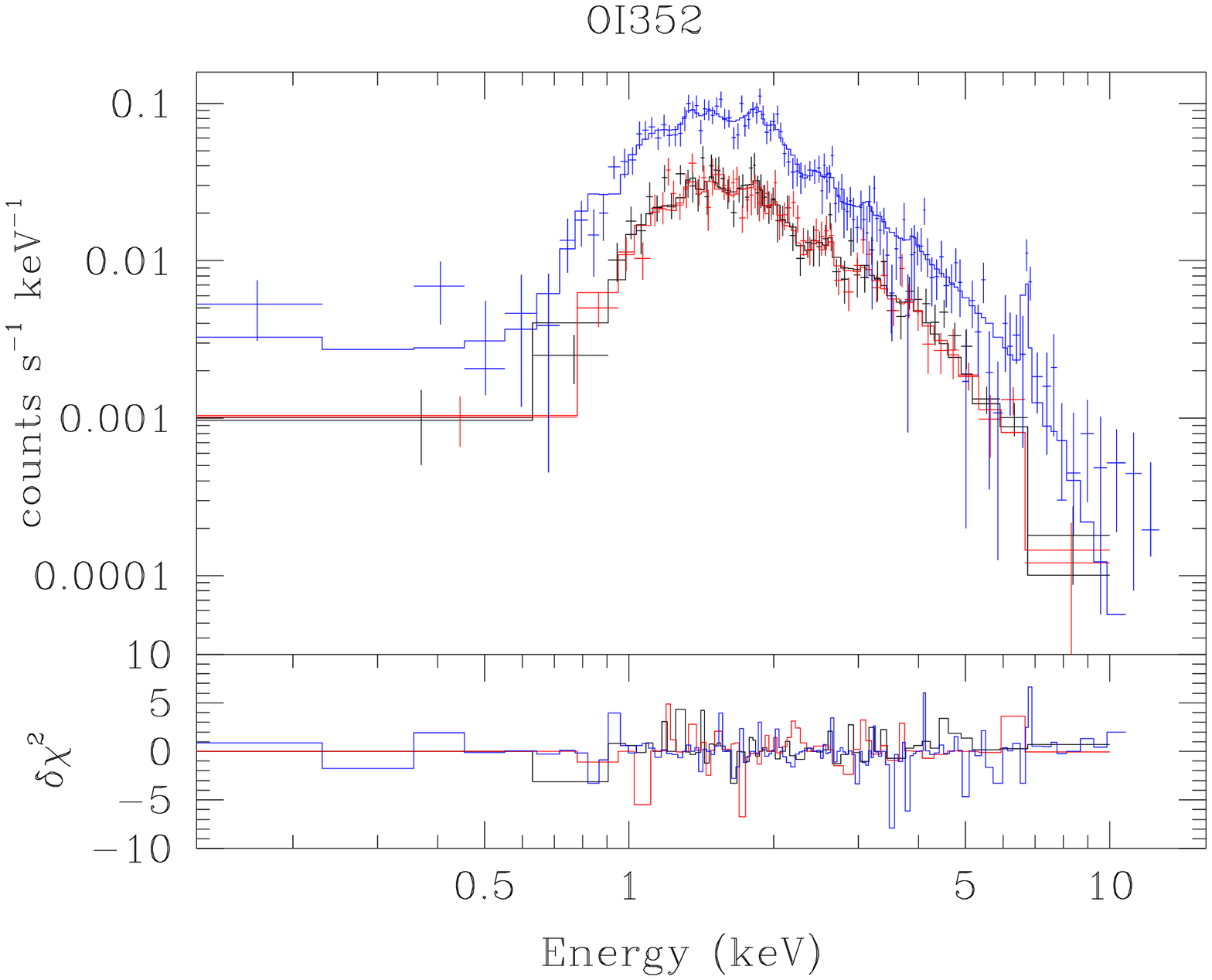}
\end{minipage}
\caption{EPIC spectra of CEN\,1a,b (left) and OI\,352 (right) along with our best fit models (see Table\,\ref{fitsO}). The lower panels illustrate the residuals in the form of contributions to the overall $\chi^2$.}
\label{specsO}
\end{figure*}

All O-type stars detected with EPIC have X-ray spectra that are best fit with a thermal plasma model. The majority of the O-type stars in our sample have intrinsically rather soft spectra with kT between 0.3 and 0.6\,keV. This is in line with the overall properties of O-type stars as derived from CCD X-ray spectra \citep{Naze} and is consistent with the general picture of intrinsic X-ray emission due to instabilities embedded in the stellar wind. There are two notable exceptions to this rule. These are CEN\,1a,b and OI\,352. The spectra of these stars are shown in Fig.\,\ref{specsO}. They both display an Fe K line and require a second, hotter, temperature of 3.8 and 2.5\,keV for CEN\,1a,b and OI\,352 respectively. It is worth comparing our best-fit spectral parameters to those found by \citet{Broos} with the ACIS-I instrument aboard {\it Chandra}. 

Concerning CEN\,1a,b, we must recall that our spectrum includes both CEN\,1a and CEN\,1b and is likely also contaminated by CEN\,37, although the latter was found to be 60 times fainter than the pair CEN\,1a,b with {\it Chandra}. These three objects were resolved by {\it Chandra} \citep[sources 536, 543 and 574 respectively in][]{Broos}. Our EPIC spectra indicate a prominent Fe K line (much stronger than in the ACIS-I spectra\footnote{We note that the ACIS-I spectrum of CEN\,1b is likely affected by pile-up.}) that constrains the temperature of the hotter plasma component quite well. This temperature (3.8\,keV) is significantly lower than the values of the hotter components found in the ACIS-I spectral fits, which were 15\,keV for CEN\,1a and 10.4\,keV for CEN\,1b \citep{Broos}. CEN\,37 was reported to have an intrinsically soft spectrum \citep[kT = 0.6\,keV,][]{Broos} and should thus not contribute to the hard emission. Temperatures above 10\,keV as advocated by \citet{Broos} are very difficult to understand within the conventional scenarios \citep[wind instabilities or magnetically channeled winds for the intrinsic emission, colliding winds in binary systems; see][for a review]{GN} for the X-ray emission of massive stars and would be reminiscent of so-called $\gamma$\,Cas-like variables. Our new result is much more in line with typical emission from a colliding wind binary system as would be expected from a Trapezium-like system such as CEN\,1a,b, displaying non-thermal radio emission \citep{RGM}. The observed X-ray flux of CEN\,1a,b in the {\it XMM-Newton} observation ($4.23 \times 10^{-12}$\,erg\,cm$^{-2}$\,s$^{-1}$) exceeds the combined value of CEN\,1a and CEN\,1b in the {\it Chandra} observation by about 37\%. This difference could result from the combination of various effects, including pile-up of the ACIS-I spectra, some differences in the adopted energy ranges (0.5 -- 8.0\,keV for {\it Chandra} versus 0.5 -- 10\,keV used here), as well as genuine orbital variability, which is a common feature among colliding wind massive binaries. 

OI\,352 is the second brightest source in our data. Its spectrum is dominated by a hard emission component (kT = 2.6\,keV). \citet{Broos} reported a temperature of 4.4\,keV for the hotter plasma component. Again, our temperature is lower, but the difference is less dramatic than for CEN\,1. There is however a rather large discrepancy in the observed fluxes between {\it Chandra} ($f_X = 4.44 \times 10^{-13}$\,erg\,cm$^{-2}$\,s$^{-1}$) and {\it XMM-Newton} ($f_X = 11.06 \times 10^{-13}$\,erg\,cm$^{-2}$\,s$^{-1}$). Such a difference could hardly be attributed to instrumental effects and would thus likely indicate genuine flux variability. The high plasma temperature, the high X-ray luminosity (for an O8 star) and the possible variability all point at this system being a colliding wind binary candidate.\\ 

\begin{figure}
\centering
\includegraphics[width=9cm]{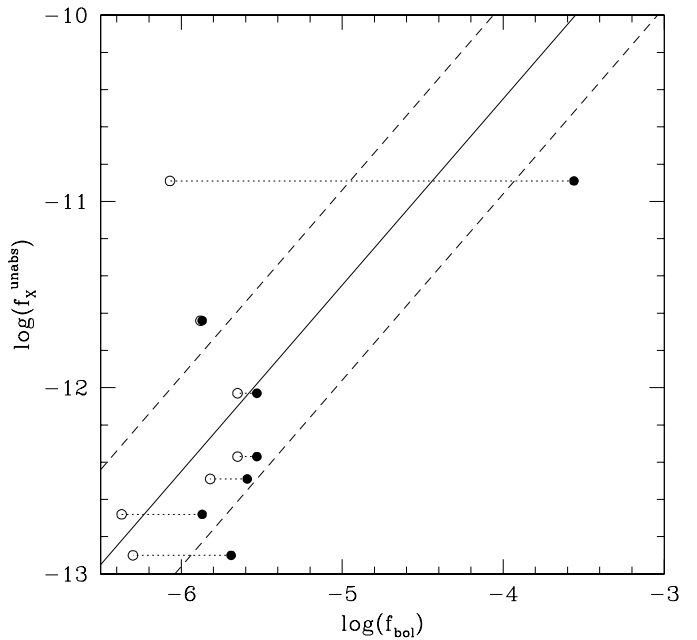}
\caption{Absorption corrected X-ray flux of the O-type stars versus bolometric flux. The filled symbols yield the bolometric fluxes derived from the $V$ magnitudes and the $A_V$ values, whilst the open symbols yield the bolometric fluxes obtained assuming the stars are typical main-sequence O-stars at a distance of 2.1\,kpc. See Sect.\,\ref{XspecO} for details. The solid line yields the $\log{(f_{\rm X}^{\rm unabs}/f_{\rm bol})} = -6.45$ relation from \citet{Naze}. The dashed lines correspond to the dispersion about this relation.}
\label{lxlbol}
\end{figure}

Using the interstellar absorption corrected X-ray fluxes and an estimate of the bolometric fluxes of the O-type stars, we could in principle infer the value of $\log{(f_{\rm X}^{\rm unabs}/f_{\rm bol})}$ of the O-type stars in M\,17 and compare it to the typical value of $-6.45 \pm 0.51$ inferred by \citet{Naze}. In the present case, the main uncertainties on $\log{(f_{\rm X}^{\rm unabs}/f_{\rm bol})}$ stem from the highly uncertain bolometric luminosities. This is illustrated in Fig.\,\ref{lxlbol}, where we have adopted two different approaches for estimating $f_{\rm bol}$. First we use the observed $V$ magnitudes, the $A_V$ values adopted for the evaluation of $N_H$, and the bolometric corrections from \citet{MartinsPlez}. This yields an extremely large value for the bolometric flux of the CEN\,1a,b system which would then be X-ray under-luminous. Whilst this method is independent of the actual distance of the stars, it is extremely sensitive to any uncertainties in $A_V$. Alternatively, we have adopted the typical bolometric luminosities from \citet{Martins}, assuming a distance modulus of 11.61 and accounting for the multiplicity of CEN\,1a, 1b and 18 \citep{Hoff} as well as for the suspected multiplicity of OI\,352. This method is hence sensitive to the uncertainty on the distance modulus as well as to the assumptions on multiplicity. Figure\,\ref{lxlbol} illustrates the huge impact of the uncertainty on the bolometric flux in the case of CEN\,1a,b which would be highly X-ray over-luminous within the second scenario. Nevertheless, we can conclude that on average and within the uncertainties described above, the X-ray luminosities of the O-stars in M\,17 follow the standard scaling relation of \citet{Naze} quite well. 

\subsubsection{Other point sources}
In addition to the O-stars, we have also analyzed the spectra of 13 relatively bright point sources in the field of M\,17. Nine of them have a near-IR counterpart (see Table\,\ref{fitsPMS}). All spectra were fitted both with thermal plasma models, adopting the same abundances as for the O-stars, and with non-thermal power-law models. The latter would be relevant for background AGN, unrelated to M\,17. 

The results of these fits are displayed in Table\,\ref{fitsPMS}. The thermal fits indicate very high temperatures, exceeding 2.4\,keV, for all sources. Furthermore, with one exception, the best-fit column densities exceed $0.5 \times 10^{22}$\,cm$^{-2}$ and reach $3.4 \times 10^{22}$\,cm$^{-2}$ for source \#33. This corresponds to $E(B-V) = 5.86$ or $A_V = 22.8$, adopting the $R_V$ value of \citet{Hoff}. Hence, it seems unlikely that our sample is significantly polluted by foreground stars and most of these sources probably either belong to M\,17 or are background objects. Assuming that these sources are located at the distance of M\,17 and that the thermal plasma models are adequate, the absorption-corrected fluxes correspond to X-ray luminosities in the range $4.6 \times 10^{31}$ to $5.5 \times 10^{32}$\,erg\,s$^{-1}$. The upper limit of this range corresponds to the highly absorbed source \#33 and is thus more uncertain because of the correction of the interstellar absorption. If we discard this object, the highest luminosity still amounts to $2.2 \times 10^{32}$\,erg\,s$^{-1}$.

\begin{table*}[t!]
\caption{Results of the fits of the EPIC spectra of non-O-star point-like sources in the field of M\,17 with models of the kind {\tt wabs*vapec} (upper part of the table) and {\tt wabs*power} (lower part). In the thermal plasma models, the abundances of the hot plasma were taken to be 1.0, 2.4, 0.4 and 0.7 times solar for He, C, Fe and all other metals, respectively (see text). Column 4 provides the reduced $\chi^2$ of the fit as well as the number of degrees of freedom (in brackets). The observed and absorption-corrected X-ray fluxes are evaluated in the 0.5 -- 10\,keV energy range.\label{fitsPMS}}
\begin{center}
\begin{tabular}{l c c c c c c c c}
\hline
XID & $N_H$ & kT & $\chi^2_{\nu}$ (d.o.f.) & $f_X$ & $f_X^{\rm unabs}$  & $J$ & $H$ & $K_s$ \\
\#  & $10^{22}$\,cm$^{-2}$ & keV &   & $10^{-13}$\,erg\,cm$^{-2}$\,s$^{-1}$ & $10^{-13}$\,erg\,cm$^{-2}$\,s$^{-1}$ & & &\\
\hline
 4 & $0.5^{+0.6}_{-0.2}$ & $2.8^{+5.0}_{-1.6}$ & 1.39 (24) & 0.58 & 0.88 & 10.62 & 9.65 & 9.12\\
24 & $1.8^{+1.1}_{-0.8}$ & $\geq 2.5$ & 1.57 (13) & 1.44 & 2.57 & 14.50 & 13.04 & 12.01 \\
33 & $3.4^{+0.9}_{-0.8}$ & $7.0^{+16.1}_{-2.8}$ & 0.95 (34) & 5.48 & 10.42 \\
37 & $2.1^{+1.0}_{-0.7}$ & $2.5^{+1.7}_{-0.7}$ & 1.14 (50) & 1.08 & 2.79  & 13.83 & 12.44 & 11.82 \\
40 & $1.0^{+0.8}_{-0.3}$ & $\geq 7.9$ & 1.65 (20) & 1.31 & 1.67 & 13.09: & 14.31 & 12.84 \\
42 & $0.3^{+0.2}_{-0.1}$ & $\geq 3.9$ & 1.27 (54) & 0.85 & 1.01 & 12.49: & 11.52: & 11.23 \\
46 & $1.7^{+0.5}_{-0.4}$ & $3.4^{+1.8}_{-0.9}$ & 0.94 (59) & 1.78 & 3.51 \\
47 & $2.2^{+1.1}_{-0.8}$ & $3.4^{+5.0}_{-1.4}$ & 1.36 (21) & 1.59 & 3.42 \\
51 & $0.7^{+0.3}_{-0.2}$ & $3.5^{+1.9}_{-1.0}$ & 0.97 (71) & 1.44 & 2.24  & 12.80 & 11.68 & 11.14\\
52 & $1.5^{+0.4}_{-0.3}$ & $2.4^{+0.7}_{-0.5}$ & 1.05 (69) & 1.83 & 4.24 & 12.65 & 11.27 & 10.64 \\
62 & $2.4^{+2.3}_{-0.9}$ & $\geq 2.6$ & 0.92 (25) & 1.37 & 2.03 \\
64 & $1.0^{+0.4}_{-0.2}$ & $7.8^{+9.2}_{-3.9}$ & 1.03 (71) & 2.96 & 4.26 & 15.07: & 13.25 & 10.98 \\
79 & $0.5^{+1.0}_{-0.2}$ & $2.7^{+2.1}_{-1.7}$ & 1.33 (30) & 0.56 & 0.87 & 11.89 & 10.76 & 10.38 \\
\hline\hline
XID & $N_H$ & $\Gamma$ & $\chi^2_{\nu}$ (d.o.f.) & $f_X$ & $f_X^{\rm unabs}$ \\
\#  & $10^{22}$\,cm$^{-2}$ & &   & $10^{-13}$\,erg\,cm$^{-2}$\,s$^{-1}$ & $10^{-13}$\,erg\,cm$^{-2}$\,s$^{-1}$ \\
\hline
24 & $2.3^{+1.5}_{-1.0}$ & $2.2^{+0.9}_{-0.7}$ & 1.52 (24) & 1.46 & 3.59 & 14.50 & 13.04 & 12.01 \\
40 & $0.7^{+0.9}_{-0.4}$ & $1.0^{+0.6}_{-0.4}$ & 1.60 (20) & 1.44 & 1.64 & 13.09: & 14.31 & 12.84 \\
47 & $2.7^{+1.5}_{-1.1}$ & $2.6^{+0.9}_{-0.8}$ & 1.35 (21) & 1.60 & 5.98 \\
62 & $3.1^{+3.1}_{-1.9}$ & $1.8^{+1.5}_{-1.1}$ & 0.91 (24) & 1.26 & 2.53 \\
64 & $1.3^{+0.4}_{-0.3}$ & $1.9^{+0.3}_{-0.3}$ & 1.00 (71) & 2.97 & 5.08 & 15.07: & 13.25 & 10.98\\
79 & $0.8^{+0.4}_{-0.3}$ & $2.6^{+0.8}_{-0.6}$ & 1.30 (30) & 0.56 & 1.34 & 11.89 & 10.76 & 10.38\\
\hline
\end{tabular}
\end{center}
\end{table*}
\normalsize
The high temperatures, the fact that most sources are located in M\,17 or behind, the range of luminosities as well as the location of the near-IR counterparts in the color-magnitude diagram  (Fig.\,\ref{2MASS}) suggest that these sources could be pre-main sequence (PMS) stars. PMS stars reach their highest X-ray luminosities during flares and we have thus inspected the X-ray light curves of all the sources of our spectroscopic sample. Whilst most of the sources do not show a significant, coherent variability between the different EPIC cameras, source \#37 undergoes a flaring event towards the end of the observation, although our data only cover the onset of the flare, with the count rate increasing by a factor $\sim 25$.

For about half of the sources in our sample, a slightly better fit quality is obtained with an absorbed power-law model. These fits are listed in the lower part of Table\,\ref{fitsPMS}. When fitted with thermal plasma models, these sources indicate rather high best-fit temperatures: in some cases, only a lower limit to kT could be obtained. 

One source that is of special interest is \#52. It is the third brightest non-O-star source with an X-ray luminosity of $2.2 \times 10^{32}$\,erg\,s$^{-1}$. This object has two {\it Chandra} counterparts, CEN\,98 \citep[photometric spectral type B4\,V,][]{CEN} and CEN\,97 \citep[photometric spectral type B1\,V,][]{CEN}, both located at 1.68 arcsec from the {\it XMM} source. There is no UV counterpart detected with the Optical Monitor, which is consistent with a deeply embedded source, as suggested by the large $N_H$ and the $A_V$ of 10.6 and 11.0\,mag quoted by \citet{CEN} for CEN\,97 and 98 respectively. \citet{Broos} derived observed X-ray fluxes of $5.0 \times 10^{-14}$ and $6.3 \times 10^{-14}$\,erg\,cm$^{-2}$\,s$^{-1}$, which combined together are slightly less than our value ($1.83 \times 10^{-13}$\,erg\,cm$^{-2}$\,s$^{-1}$). This source therefore apparently corresponds to a pair of early-mid B-type stars with an exceptionally large X-ray luminosity. Indeed, typical X-ray luminosities of early B main sequence stars are about a factor ten below the result found here \citep{CCCPN} and many of the mid-late B stars detected in X-rays are likely binaries with a low-mass PMS companion being the X-ray emitter \citep{CCCPE}. It would be highly interesting to re-investigate the spectral types of these stars and to look for multiplicity or other peculiarities. 

\subsection{The diffuse emission}
As stated in Sect.\,\ref{intro}, soft diffuse X-ray emission from the M\,17 superbubble was detected with previous X-ray observatories \citep{Dunne,Townsley1,Hyodo}. This emission is strongest around the core of the ionizing cluster, NGC\,6618, and further extends towards the east-south-east direction as an elongated structure \citep[][see also Fig.\,\ref{X3col}]{Townsley2}. 

\citet{Dunne} noted a huge discrepancy between the observed level of X-ray emission and the two orders of magnitude higher level expected for a model of a wind-blown superbubble including heat conduction \citep[][ see also A\~{n}orve-Zeferino et al.\ \citeyear{Anorve}]{Weaver}. They suggested that this discrepancy stems from the effect of heat conduction and found a better agreement for models where heat conduction is suppressed (probably by magnetic fields), but dynamical mixing of cold interstellar material with the hot gas of the bubble is included.

Because of their superb angular resolution, the {\it Chandra} data rule out the possibility that a substantial fraction (more than about 8\%) of the diffuse emission could be due to unresolved late-type pre-main sequence stars \citep{Townsley2}. These data yielded an (absorption corrected) X-ray luminosity of $L_X = 3.4 \times 10^{33}$\,erg\,s$^{-1}$ in the 0.5 -- 2\,keV band assuming a distance of 1.6\,kpc \citep{Townsley2}. A very similar luminosity ($L_X = 3.5 \times 10^{33}$\,erg\,s$^{-1}$ in the 1.0 -- 8.0\,keV energy band) was obtained by \citet{Hyodo} from a {\it Suzaku} observation.  

\citet{Reyes} presented 3-D hydrodynamical simulations of the M\,17 superbubble and its X-ray emission. They successfully reproduced the observed champagne flow morphology and luminosity of the X-ray emission. These authors demonstrate the importance of the inhomogeneous interstellar medium and of the interaction of the cluster wind with the dense molecular cloud for correctly reproducing the observed X-ray luminosity.\\

We have first extracted the spectrum of the diffuse emission around the core of NGC\,6618, excluding all detected point sources. This spectrum can be fitted with a two-temperature thermal plasma model with kT of 0.8 and $\sim 4.0$\,keV \citep{Mernier}. The higher temperature is most probably an indication of residual contamination by a population of unresolved point sources associated with pre-main sequence stars. 

\begin{figure}
\centering
\includegraphics[width=9cm]{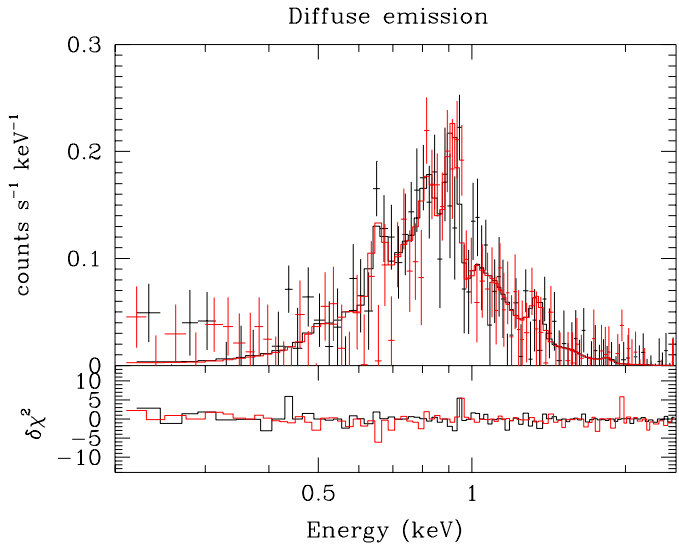}
\caption{MOS\,1 and MOS\,2 spectra of the diffuse X-ray emission in M\,17.}
\label{M17diffuse}
\end{figure}

As a next step, we have then extracted the diffuse X-ray emission within a rectangular box centered on (18:20:52.1, $-$16:12:09), with a size of $8 \times 2.2$\,arcmin$²$ and inclined by $20^{\circ}$ with respect to the west-east direction, excluding two point sources. This box is slightly smaller than the $8 \times 3.5$\,arcmin$^2$ region used by \citet{Townsley2} and avoids the central part of NGC\,6618 whilst extending somewhat farther away from the cluster. The background was extracted over a circular area of radius 0.9\,arcmin centered on (18:20:33.1, $-$16:13:44). The resulting EPIC-MOS spectra are illustrated in Fig.\,\ref{M17diffuse}. They are qualitatively similar to the spectrum presented by \citet{Townsley2}. The latter authors used a two-temperature model with one component (kT = 0.13\,keV) with solar composition and another one (kT = 0.6\,keV) with non-solar abundances. The {\it Suzaku} observation was fitted with a single temperature (kT = 0.25\,keV) plasma model with clearly sub-solar metallicity \citep{Hyodo}. 

Here, we have adopted a different approach, fixing the plasma abundances of all components to the same mix as used for the point sources in M\,17. The best-fit ($\chi^2_{\nu} = 0.89$ for 169 d.o.f.) parameters are $N_H = 0.76^{+.13}_{-.19} \times 10^{22}$\,cm$^{-2}$, kT$_1$ = $0.11^{+.95}_{-.11}$\,keV, kT$_2$ = $0.22^{+.12}_{-.03}$\,keV, EM$_2$/EM$_1$ = 0.20 and $f_X = 5.76 \times 10^{-13}$\,erg\,cm$^{-2}$\,s$^{-1}$. The observed flux is about half the value found with {\it Chandra}. This difference most likely reflects the somewhat different extraction regions used in the present study and in the analysis of the {\it Chandra} data. Our study thus reveals that the spectral parameters derived from the spectrum of the diffuse emission are quite sensitive to the treatment of several aspects, such as the plasma abundances, the definition of the extraction regions of the source and the background, and the possible contamination by unresolved point sources. 

\section{Summary and conclusions \label{conclusions}}
In this paper, we have analyzed X-ray and UV data obtained during a 30\,ksec {\it XMM-Newton} observation of M\,17. 

The brightest X-ray point sources are associated with seven O-type stars or systems of such stars. The X-ray images are dominated by the Trapezium-like system CEN\,1a,b ((O4\,V + O) + (O4\,V + O)) and the O8\,V star OI\,352. In both cases we have shown that the EPIC spectra yield plasma temperatures that are significantly lower than found in a previous study by \citet{Broos} and are in very good agreement with a colliding winds scenario. An optical or near-IR spectroscopic monitoring of these stars, although challenging because of the large interstellar absorption and unknown orbital periods, would be extremely useful to establish their multiplicity and provide their orbital properties. 

There is a very large uncertainty on the $L_{\rm X}/L_{\rm bol}$ ratios of the O-type stars in M\,17. This uncertainty mainly stems from the impact of the huge (and sometimes poorly known) reddening on the estimates of the bolometric luminosities. Whilst the $L_{\rm X}/L_{\rm bol}$ ratios agree on average with the canonical relation established for a large sample of O-stars, the large uncertainties in the present case prevent us from a more detailed comparison that might reveal subtle differences and environmental effects.  

The majority of the secondary X-ray sources are likely low-mass pre-main sequence stars, with about 1/6 of the objects showing some evidence of near-IR excesses attributable to circumstellar material. The near-IR counterparts of our X-ray sources span a wide range in reddening, which makes a quantitative comparison of their colors and magnitudes with theoretical evolutionary tracks very difficult. Out of the 13 brightest X-ray sources, not associated with an O-star, only one displays a flare in our 30\,ksec observation. This can be compared to the results of the 40\,ksec {\it Chandra} observation: \citet{Broos} reported significant variability for 39 out of 886 sources, with a wide range of flare morphologies. Beside this short-term (flaring) activity, there must also be a substantial long-term (of order months or years) variability of the X-ray sources of M\,17. Indeed, the relation between the ACIS-I count rates, recorded in March 2002, and the EPIC-MOS rates, measured twenty months later, exhibits a very important scatter which likely reflects this variability. 

There is little correspondence between the 92 X-ray sources and the 614 UV detections. Most of the UV sources are in fact foreground objects, unrelated to M\,17. The UV data also reveal some nebular emission eastwards of the central cluster. 

Throughout this work, we have fitted the X-ray spectra of point and diffuse sources with plasma models accounting for the specific metal abundances of the M\,17 nebula as determined by \citet{Garcia}, except for the iron abundance. Indeed, the latter is clearly higher in our spectra than reported by \citet{Garcia}. This is an interesting puzzle and, although an in-depth analysis of this effect is beyond the scope of the present paper, we suggest that this discrepancy might be related to ionization effects. The presence of a very hot plasma (as revealed by the diffuse X-ray emission) might impact on the ionization stages of iron and could bias the abundances inferred from optical diagnostics ([Fe\,{\sc ii}] and [Fe\,{\sc iii}] lines).


\section*{Acknowledgements}
We thank Dr.\ Ya\"el Naz\'e for sharing her spectral extraction routines with us. We acknowledge support from the Fonds de Recherche Scientifique (FRS/FNRS), through the XMM/INTEGRAL PRODEX contract as well as by the Communaut\'e Fran\c caise de Belgique - Action de recherche concert\'ee - Acad\'emie Wallonie - Europe. This research made use of data products from the Two Micron All Sky Survey, which is a joint project of the University of Massachusetts and the Infrared Processing and Analysis Center/California Institute of Technology, funded by NASA and NSF.



\bibliographystyle{elsarticle-harv}







\end{document}